\newif\ifANON{}
\newif\ifACM{}
\newif\ifLNCS{}
\newif\ifIACRTRANS{}
\newif\ifIEEE{}
\newif\ifUSENIX{}
\newif\ifLINUXBUILD{}

\LNCStrue{}

\IfFileExists{/dev/null}{\LINUXBUILDtrue}{}

\ifACM{}
\ifANON{}
\documentclass[sigconf,review=true,anonymous=true,nonacm=true]{acmart}
\else
\documentclass[sigconf,nonacm=true]{acmart}
\fi
\author{Cesar Pereida Garc\'ia}
\orcid{0000-0001-6812-8498}
\affiliation{
\institution{Tampere University}
\city{Tampere}
\country{Finland}
}
\email{cesar.pereidagarcia@tuni.fi}

\author{Billy Bob Brumley}
\orcid{0000-0001-9160-0463}
\affiliation{
\institution{Tampere University}
\city{Tampere}
\country{Finland}
}
\email{billy.brumley@tuni.fi}
\fi

\ifLNCS{}
\documentclass[runningheads]{llncs}
\usepackage{xcolor}
\usepackage[colorlinks,pagebackref,breaklinks]{hyperref}
\hypersetup{
  hypertexnames=true,
  linkcolor=blue,
  anchorcolor=black,
  citecolor=blue,
  urlcolor=magenta
}

\usepackage{color, colortbl}
\definecolor{Gray}{gray}{0.9}

\usepackage{pifont}
\newcommand{\cmark}{\ding{51}}%
\newcommand{\xmark}{\ding{55}}%

\usepackage[square,comma,numbers,sectionbib]{natbib}

\makeatletter
\renewcommand\@biblabel[1]{#1.}
\makeatother
\ifANON\else
\author{Iaroslav Gridin\orcidID{0000-0002-1239-1841}
\institute{Tampere University, Tampere, Finland}
\and
Antonis Michalas\orcidID{0000-0002-0189-3520}
\institute{Tampere University, Tampere, Finland}
}
\fi
\fi

\ifIACRTRANS{}
\ifANON{}
\documentclass[submission]{iacrtrans}
\else
\documentclass[final]{iacrtrans}
\fi
\usepackage[backend=bibtex,style=alphabetic,maxbibnames=99,natbib=true]{biblatex}

\author{Iaroslav Gridin}

\authorrunning{Iaroslav Gridin}%
\index{Gridin, Iaroslav}%

\institute{Tampere University, Tampere, Finland\\
\email[iaroslav.gridin@tuni.fi]{{iaroslav.gridin}@tuni.fi}
}

\bibliography{manuscript_rw,manuscript_ro}
\fi

\ifIEEE{}
\documentclass[10pt,conference,compsocconf]{IEEEtran}
\usepackage[square,sort&compress,numbers]{natbib}

\usepackage[colorlinks,pagebackref,breaklinks]{hyperref}
\hypersetup{
  hypertexnames=true,
  linkcolor=blue,
  anchorcolor=black,
  citecolor=blue,
  urlcolor=blue
}
\ifANON\else
\author{\IEEEauthorblockN{Iaroslav Gridin}}
\IEEEauthorblockA{Tampere University, Tampere, Finland}
\fi
\fi

\ifUSENIX{}
\documentclass[letterpaper,twocolumn,10pt]{article}
\makeatletter
\@namedef{ver@breakurl.sty}{}
\@namedef{ver@cite.sty}{}
\makeatother
\usepackage[hyphens]{url} 
\usepackage{usenix}
\date{}
\ifANON\else
\author{
\textnormal{Iaroslav~Gridin}\\
Tampere University, Tampere, Finland
} 

\fi
\fi

\usepackage[T1]{fontenc}
\usepackage{graphicx}
\usepackage{xspace}
\usepackage{lipsum}
\usepackage[frozencache]{minted}
\usepackage{array}
\usepackage[shortlabels]{enumitem}
\usepackage[shortlabels]{enumitem}
\usepackage{booktabs}

\ifLNCS{}

\fi
\ifIACRTRANS{}

\fi
\ifUSENIX{}

\fi
\ifACM{}

\fi
\ifIEEE{}
\makeatletter
\renewcommand{\@IEEEsectpunct}{~}
\makeatother

\fi

\newcommand{\code}[1]{\texttt{#1}}

\ifANON{}
\def\acvprust{\textsc{ACVPFrame}}
\else
\def\acvprust{\textsc{acvp-rust}}
\fi

\newcommand{\KEYWORDS}{%
ACVP\null;
Coverage\null;
Cryptography\null;
Fuzzing\null;
KAT\null;
NSS\null;
Testing;
}

\title{Point Intervention: Improving ACVP Test Vector Generation Through Human Assisted Fuzzing\thanks{\protect\footnotesize This work was partially funded by the EU research project SWARMCHESTRATE (No. 101135012) and the Mozilla Corporation.}}

\ifACM{}
\begin{abstract}
{Automated Cryptographic Validation Protocol} ({ACVP}) is an existing protocol that is used to validate a software or hardware cryptographic module automatically. In this work, we present a system providing the method and tools to produce well-covering tests in {ACVP} format for cryptographic libraries. The system achieves better coverage than existing fuzzing methods by using a hybrid approach to fuzzing cryptographic primitives. In addition, the system offers a framework that allows to creates easily and securely create testing modules for cryptographic libraries. The work demonstrates how this system has been used to improve automated testing of {NSS} (Network Security Services), a popular cryptographic library, detect its vulnerabilities and suggest ways to improve and further develop the ACVP test format.

\end{abstract}
\keywords{\KEYWORDS{}}
\fi

\begin{document}

\titlerunning{\acvprust}
\maketitle

\ifLNCS{}
\begin{abstract}
{Automated Cryptographic Validation Protocol} ({ACVP}) is an existing protocol that is used to validate a software or hardware cryptographic module automatically. In this work, we present a system providing the method and tools to produce well-covering tests in {ACVP} format for cryptographic libraries. The system achieves better coverage than existing fuzzing methods by using a hybrid approach to fuzzing cryptographic primitives. In addition, the system offers a framework that allows to creates easily and securely create testing modules for cryptographic libraries. The work demonstrates how this system has been used to improve automated testing of {NSS} (Network Security Services), a popular cryptographic library, detect its vulnerabilities and suggest ways to improve and further develop the ACVP test format.

\keywords{\KEYWORDS{}}
\end{abstract}
\fi

\ifIACRTRANS{}
\begin{abstract}
{Automated Cryptographic Validation Protocol} ({ACVP}) is an existing protocol that is used to validate a software or hardware cryptographic module automatically. In this work, we present a system providing the method and tools to produce well-covering tests in {ACVP} format for cryptographic libraries. The system achieves better coverage than existing fuzzing methods by using a hybrid approach to fuzzing cryptographic primitives. In addition, the system offers a framework that allows to creates easily and securely create testing modules for cryptographic libraries. The work demonstrates how this system has been used to improve automated testing of {NSS} (Network Security Services), a popular cryptographic library, detect its vulnerabilities and suggest ways to improve and further develop the ACVP test format.

\keywords{\KEYWORDS{}}
\end{abstract}
\fi

\ifIEEE{}
\begin{abstract}
{Automated Cryptographic Validation Protocol} ({ACVP}) is an existing protocol that is used to validate a software or hardware cryptographic module automatically. In this work, we present a system providing the method and tools to produce well-covering tests in {ACVP} format for cryptographic libraries. The system achieves better coverage than existing fuzzing methods by using a hybrid approach to fuzzing cryptographic primitives. In addition, the system offers a framework that allows to creates easily and securely create testing modules for cryptographic libraries. The work demonstrates how this system has been used to improve automated testing of {NSS} (Network Security Services), a popular cryptographic library, detect its vulnerabilities and suggest ways to improve and further develop the ACVP test format.

\end{abstract}
\fi

\ifUSENIX{}
\begin{abstract}
{Automated Cryptographic Validation Protocol} ({ACVP}) is an existing protocol that is used to validate a software or hardware cryptographic module automatically. In this work, we present a system providing the method and tools to produce well-covering tests in {ACVP} format for cryptographic libraries. The system achieves better coverage than existing fuzzing methods by using a hybrid approach to fuzzing cryptographic primitives. In addition, the system offers a framework that allows to creates easily and securely create testing modules for cryptographic libraries. The work demonstrates how this system has been used to improve automated testing of {NSS} (Network Security Services), a popular cryptographic library, detect its vulnerabilities and suggest ways to improve and further develop the ACVP test format.

\end{abstract}
\fi



\section{Introduction}\label{sec:intro}

Testing computer software is a sine qua non that ensures proper functionality.
Numerous implementation issues arise due to human errors.
A typical example lays in programs lacking features that check input size in order to prevent access attempts after the end of an array.
While modern programming languages offer various mechanisms to mitigate
issues, such as
advanced type systems, performance is paramount in writing cryptographic software. As a result, these programs often rely on direct memory access and are typically written in languages like C or C++~\cite{rustytypes}.

One method of ensuring low-level code correctness is external automated testing.
Automated testing is a process that verifies the execution of a program without human interaction, thus significantly reducing costs.
Typically, testing involves issuing challenges to the program and validating responses.
For example, a test might involve ``encrypting bytes A with key \(\mathsf{K}\) and verifying that the output matches bytes B''.
Tests can be generated on demand or pre-generated, and may verify results against pre-existing values or another program, or simply confirm that the program is executed without errors.
A crucial aspect of testing is \textit{coverage}, which measures how thoroughly the code is tested to ensure that no portion remains untested.

Often, well-covering test sets are produced by fuzz testing, or fuzzing for short. Fuzzing is a form of automatic testing, which repeatedly runs the target software with mutated input.
In recent years, coverage-based grey-box fuzzing (CGF) has emerged as an effective way of locating security issues~\cite{weizz}.
CGF involves instrumenting the code by inserting special markers to collect coverage data.
It then utilizes changes in coverage as a guide to identify areas of input to be modified in order to maximize coverage and gain insights into the structure of the input.
However, satisfying complex coverage criteria through random mutation can be resource-intensive.
To address this challenge, various additional approaches have been explored, such as leveraging hardware support~\cite{hwfuzz} and employing symbolic execution~\cite{symfuzz}.

\subsection{Automated Cryptographic Validation Protocol (ACVP)}\label{subsec:ACVP}
On July 17, 1995, NIST established the Cryptographic Algorithm Validation Program (CAVP) and the Cryptographic Module Validation Program (CMVP)
in order to validate cryptographic modules~\cite{acvt}.
Originally, all CMVP communications and operations on submitted cryptographic modules took place exclusively in testing laboratories.
However, as technology advanced, the industry demanded faster testing cycles than that scheme could provide, the required human involvement resulted in mistakes,
while modules could not be monitored after initial validation.
The Automated Cryptographic Validation Testing (ACVT) project was implemented to reduce the costs and time of validation, while still providing a high level of assurance.
As part of the ACVT project, NIST has designed and developed the Automated Cryptographic Validation Protocol ({ACVP})~\cite{acvp} -- a protocol and software for automated algorithm testing.
NIST has published specifications of the protocol~\cite{acvp} and the source code for the server~\cite{acvp-server}, while it runs both demo and production servers for remote verification.
{ACVP} is a protocol automatically testing software or hardware cryptographic modules~\cite{acvp-spec}. {ACVP} is developed by NIST and includes a portable, human-understandable, universal format of test data based on JSON~\cite{json}.
ACVP software often is categorized as one of three parties: a server, a proxy, and a client.

\begin{enumerate}
    \item The server side manages various requests, including those for test vectors and validation.
    \item A proxy equipped with ACVP enables
    communication with offline systems and facilitates the transfer of information from the tested system
    to the server and back. Sometimes, software combines functions of a proxy and a client.
    \item The client component is particularly relevant to users seeking validation for library. An ACVP client is directly connected to the module undergoing testing and communicates with the ACVP server to request test vectors, output the results of test executions, and seek algorithm validation.
\end{enumerate}

\subsection{ACVP Tests}\label{subsec:acvptests}
ACVP supports many primitives by way of \texttt{``subspecifications''}, which describe a family of cryptographic primitives like \texttt{``Secure Hash''}~\cite{acvp-sha}.
{ACVP} tests do not have a shared defined structure, but, as a rule, subspecifications describe similar layouts.
Tests are distributed in form of ``vector sets''.
Vector sets contain shared information like the algorithm and the subspecification revision, and an array of ``test groups''.
Test groups, similarly, include shared information specific to the subspecification, and an array of ``tests''.
Tests include the rest of the information.
The cryptographic module being tested has to respond to a test vector set with ``test vector response'', which is structured in a similar way.
An example of an ACVP vector set can be seen in \autoref{listing:acvp-example}.

\begin{figure}[!ht]
\begin{minted}{json}
{
  "vsId": 805548,
  "algorithm": "ACVP-AES-GCM",
  "revision": "1.0",
  "isSample": true,
  "testGroups": [{
    "tgId": 1,
    "testType": "AFT",
    "direction": "encrypt",
    "keyLen": 128,
    "ivLen": 96,
    "ivGen": "external",
    "ivGenMode": "8.2.1",
    "payloadLen": 256,
    "aadLen": 120,
    "tagLen": 104,
    "tests": [{
      "tcId": 1,
      "pt": "28E3FB...9809",
      "key": "C19A...AD2",
      "aad": "E9FB...1B",
      "iv": "C4...DEFB"
     }]
  }]
}
\end{minted}
\caption{Example of an ACVP test vector set, obtained from ACVP demo server.}%
\label{listing:acvp-example}
\end{figure}

\subsection{Contributions}\label{subsec:Contr}
The core contribution of this work lies in the development of \acvprust{} -- a comprehensive system designed to generate tests for cryptographic libraries.
This system features a human-readable, flexible, and universal format, facilitating seamless integration into existing workflows.
Several tools
interface with the ACVP (e.g.\ \code{acvpproxy}~\cite{acvpparser}) or
  work with cryptographic libraries to run vector sets (e.g.\ \code{acvpparser}~\cite{acvpparser}) or even support both (see \code{libacvp}~\cite{libacvp}). However, these tools are predominantly coded in C, posing challenges in terms of extensibility and complexity. Given the need for precise handling of ACVP tests and seamless integration with complementary tools for program execution analysis, we opted to develop our own library in \textbf{Rust}. \textbf{Rust} is renowned for its strong typing and security-focused design, 
  hence aligns seamlessly with our objectives, ensuring robustness and efficiency in our implementation efforts.

\smallskip
The core contributions of the paper can be summarized as follows:

\begin{enumerate}[\bfseries C1.]
\item Development of a software framework for producing and running test vector sets tailored for cryptographic libraries.
\item Introduction of a methodology leveraging human assistance to enhance the framework's capability in generating comprehensive test vectors.
\item Proposal of two enhancements to augment the {ACVP} test vector format, along with the introduction of novel subspecifications for {ACVP}.
\item
Completion of extensive experiments that allowed us to trace undiscovered bugs in Mozilla's {NSS} cryptographic library\footnote{\href{https://firefox-source-docs.mozilla.org/security/nss/index.html}{https://firefox-source-docs.mozilla.org/security/nss/index.html}}.
This serves as proof that the framework we designed and developed facilitates the detection of undiscovered bugs.
\end{enumerate}

\subsection{Organization}\label{subsec:Organization}

The rest of the paper is organized as follows:
\autoref{sec:Related} introduces the key fuzzing tools that closely align with our research objectives.
\autoref{sec:acvp_rust} provides an overview of \acvprust{}, detailing its architecture and design decisions.
\autoref{sec:bugs} illustrates the discovery of bugs in the cryptographic library NSS through the utilization of \acvprust{} while \autoref{sec:AnalysisCover} emphasizes in its ability to achieve enhanced code coverage.
In \autoref{sec:acvp} we assess the ACVP system and its testing format, offering suggestions for enhancements.
Finally, \autoref{sec:conclusion} concludes the paper and outlines potential future research directions to further develop \acvprust.

\section{Related Work}\label{sec:Related}
Fuzzing is a constantly developing field. Several competing mature coverage-guided fuzzers are being improved and multiple
projects increase the speed and quality of fuzzing in specific areas or conditions. Here are some examples of popular coverage-guided fuzzers and recent novel fuzzing techniques.

AFL++~\cite{10.5555/3488877.3488887} is a community-driven open-source tool that performs fuzzing.
AFL++ has been created based on patches of the original AFL
that were unmaintained for 18 months, though still popular
popular among researchers.
AFL++'s fuzzing is coverage-guided: it receives feedback from code
executed to mutate the input.
Similar to libFuzzer~\cite{libfuzzer}, AFL++ features ``Custom Mutator API''\footnote{\href{https://aflplus.plus/docs/custom\_mutators/}{https://aflplus.plus/docs/custom\_mutators/}} which allows users to supply own functions
modifying the input within given limitations, to bypass early failure points.
AFL++ uses many sophisticated methods to automatically discover new code paths, some of which are listed in the referenced paper.
AFL++ automatically produces good coverage, but will still fail to produce deep coverage often when applied to cryptographic software, due to its random nature and complexity of conditions used in cryptography.  As shown in \autoref{sec:bugs}, \acvprust{} provides an improvement over a greybox fuzzer through hybrid fuzzing, the resulting fuzzer is able to proceed through typical roadblocks.

LibFuzzer~\cite{libfuzzer} is a fuzzing tool integrated into the LLVM compiler infrastructure. LLVM~\cite{llvm} is a widely used compiler framework for several languages, which includes debugging, profiling, and other related tools.
LibFuzzer is a coverage-guided fuzzer, using LLVM to inspect running code and measure coverage.
LibFuzzer can perform structure-aware fuzzing, allowing users to supply a ``mutator''
that ensures the output has a specific structure.
LibFuzzer can interact with other LLVM tools, like sanitizers
that help discover issues
such as memory management mistakes in running code.
LibFuzzer can produce a well-covering corpus of outputs, similar to AFL++, according to tests ran by FuzzBench project~\cite{fuzzbench}, but as other fuzzers, it struggles with complex roadblocks, which are unlikely to be solved by random output generation. In this paper, we build on top of the fully automatic fuzzer to provide a framework in order to augment its output with human input: roadblocks which are by their nature difficult for a fuzzer to overcome are identified and solved by the human operator.

Fuzztruction~\cite{fuzztruction}
presents a way
of generating better outputs by mutating the program that normally produces this format. This allows us to reuse
already written code that generates the structure.
Thus the resulting fuzzer outperforms normal coverage-guided fuzzers like AFL++.
However, there is a need for the producing program as random modification of its logic has its limits.
Our work develops independently
of the available software type and relies on the interactive adjustment of structure to meet the roadblocks instead of automatic random modification of the producer.

Carpetfuzz~\cite{carpetfuzz}
uses natural language processing to extract relationships between program options from documentation.
This data is then processed into inputs
likely to elicit abnormal behavior from the program.
This approach to fuzzing is novel and has helped uncover multiple bugs,
though it relies on natural language documentation being present and covering the options we are interested in.
Our work does not rely on anything but the code itself and covers different use cases. Additionally, it is not restricted to command line options.

\section{Automatic Test Generation Framework}\label{sec:acvp_rust}

{ACVP} includes a portable and universal test format. However, there is still a need for software that allows to \textit{quickly}, \textit{easily}, and \textit{reliably} adapt it to cryptographic libraries.
We introduce \acvprust{}, a framework for producing and running ACVP test vectors.
This framework can generate test vectors with fuzzing, using code coverage feedback from cryptographic libraries, or run test vectors to validate these libraries.
We designed \acvprust{} to be modular and extensible in order to facilitate the addition of ACVP subspecifications, cryptographic library modules and instrumentations, while keeping the resulting code maintainable.

\subsection{Architecture}

\begin{figure}[!h]
  \centering
  \includegraphics[width=\columnwidth]{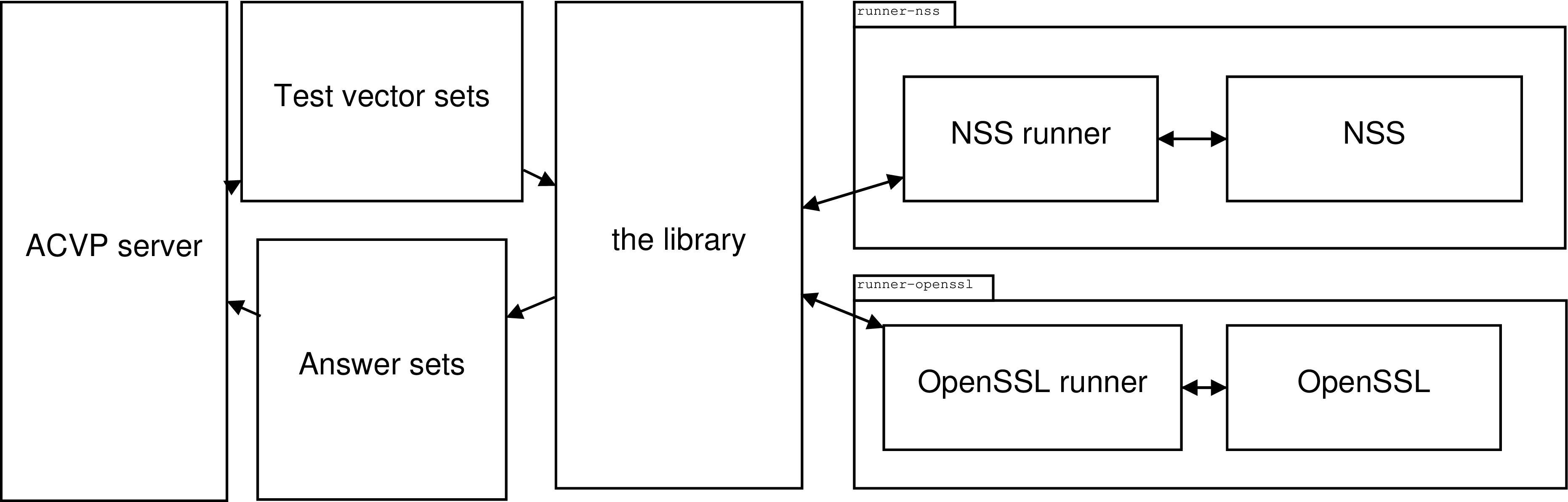}
  \caption{Structure of \acvprust{}}
\end{figure}

\acvprust{} consists of two main parts, the ``runners'' and the ``library''. ``Runners'' are adaptors that incapsulate third-party libraries or other cryptographic modules under inspection.
These provide a common interface and can be used in any combination to produce or run test vectors. ``Library'' is the shared logic that parses ACVP tests and handles the runners.
Runner and library are different processes, thus their execution is independent and library can handle any kind of unexpected behavior from a runner.
Using \acvprust{}, users can execute test vectors on a runner to validate the module or fuzz the runner's library to generate a well-covering test vector and check the cryptographic library for memory issues, crashes, or other unexpected behavior.

As a result, \acvprust{} can fuzz and instrument any library that can be compiled by {LLVM} that supports many modern languages, without much adaptation.
During fuzzing, memory handling can be checked by some of the sanitizers supported by {LLVM}.
``Library'' implements multiple ACVP subspecifications, contains tools to easily implement more, and routines for shared functionality required for related tasks.
Integration of {libFuzzer} into \textbf{Rust} ecosystem is provided by {cargo-fuzz} project~\cite{cargofuzz} that facilitates fuzzing \textbf{Rust} code or any other code linked to \textbf{Rust}.

{LibFuzzer} can be combined with multiple sanitizers, \textit{i.e.} tools that instrument the code to detect possible issues.
During our fuzzing, we used ASAN sanitizer~\cite{asan} which can detect improper memory handling while being compatible with most code.

\subsection{Hybrid Fuzzing}

\begin{figure}[!h]
  \centering
  \includegraphics[width=\columnwidth]{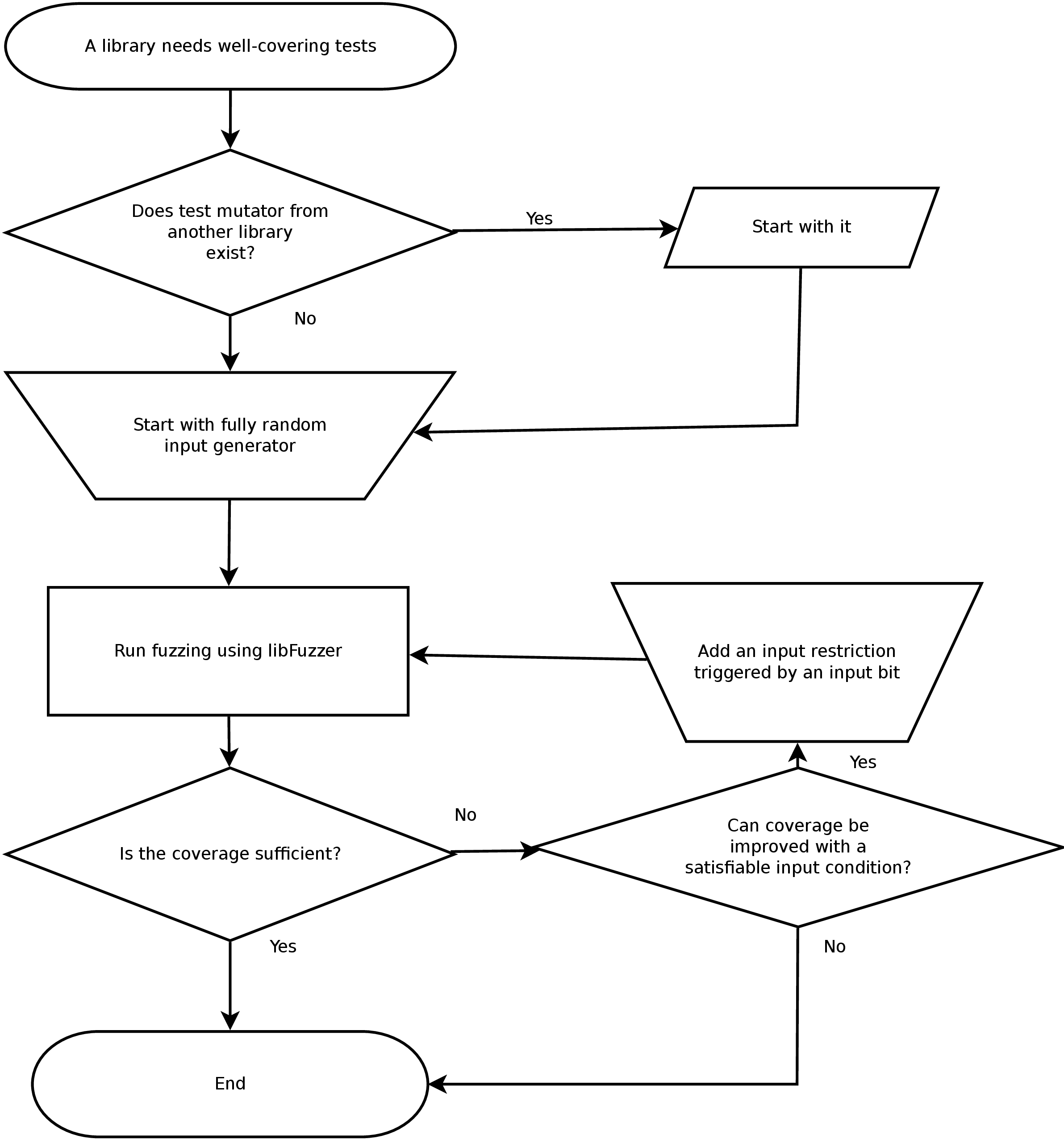}
  \caption{Flowchart of the hybrid fuzzing process.}
\end{figure}

Fuzzing tends to uncover many errors caused by an unexpected input, but when applied to cryptography or other kinds of highly structured input, it has difficulty producing deep-reaching coverage, as most inputs get discarded early.
To help with this, in \acvprust{} we use hybrid fuzzing. The method includes a simple bytestring-mutating fuzzer {libFuzzer} augmented with a domain-specific test case
translator using bits mutated by the fuzzer to decide whether to produce restricted inputs that satisfy specific conditions in the code. {LibFuzzer} can learn based on increasing code coverage and include the mutations that provide the increase.
However with specific conditions in cryptographic protocols it can take a while for the fuzzer to randomly produce an input matching them
Therefore, \acvprust{} test generators introduce special restrictions based on bit flips, like this.

\begin{minted}{rust}
let salt_len = if Arbitrary::arbitrary(u)? {
  u64::arbitrary(u)? % hash_alg.digest_len()
} else {
  u64::arbitrary(u)?
};
\end{minted}

Here, the human operator added a condition based on bit taken from randomly-mutated data to restrict the salt length to digest length, helping to avoid the input failing an early check in the library code. The coverage produced by the fuzzer indicates what is needed for the fuzzer to increase its coverage, while a related constraint can be introduced to the test case generator (for example see \autoref{fig:CoverageIncrease}).

\begin{figure}[!h]
  \centering
  \includegraphics[width=\columnwidth]{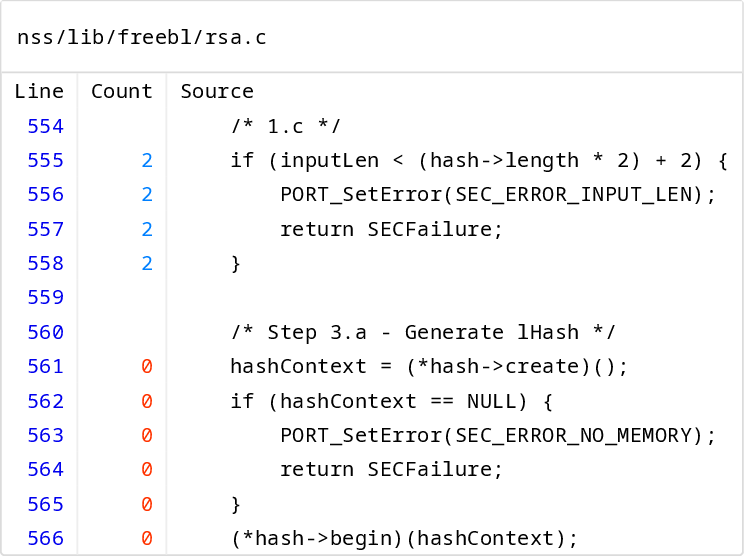}
  \caption{Example of an opportunity to add a code constraint: fuzzer fails to satisfy a condition}
\end{figure}\label{fig:CoverageIncrease}

The resulting approach combines both manual testing and fuzzing strengths: the fuzzer can automatically find deep-reaching inputs wherever possible, while manual intervention helps with the demanding parts.
Then the resulting test vector set can be used for another library. With this as a good starting point it may be extended to cover the library's special cases.
Thus, a test generator developed with one library will be useful to provide better coverage for other libraries as they are likely to need similar restrictions.
Additionally, unlike tests focused on exclusively verifying the correctness of an algorithm implementation itself, tests generated by \acvprust{} also protect against typical implementation issues by ensuring the library gracefully handles unusual or invalid input without causing security or stability issues.

\subsection{Flexibility}

The tools from \acvprust{} can be used to implement hybrid fuzzing for any library in a language compiled by {LLVM} tools. At the moment, that includes, most notably, {C}, {C++}, and \textbf{Rust}.
In \autoref{sec:bugs}, we describe how it was used to produce test coverage and find bugs in {NSS}, but the same approach can be used for any library. Tests created by the system are saved in {ACVP} standard format and can be easily examined by a human, modified, and used in any system implementing the {ACVP} specifications.

\subsection{Open Science \& Reproducible Research}
To support open science and reproducible research, and provide other researchers with the opportunity to use, test, and hopefully extend our tool, we have
\ifANON{}
anonymized our source code and made it available online\footnote{\href{https://anonymous.4open.science/r/acvpframe/}{\acvprust's Anonymized Repository: https://anonymous.4open.science/r/acvpframe/}}.
\else{}
made source code publicly available online\footnote{\href{https://gitlab.com/nisec/acvp-rust}{\acvprust's Source Code Repository: https://gitlab.com/nisec/acvp-rust}} under MPL (Mozilla Public License) 2.0.

\section{Detecting Bugs in NSS Through \acvprust{}}\label{sec:bugs}

In this section, we describe a series of bugs that we managed to uncover, while improving 
the fuzzing coverage with \acvprust. The material presented in this section can be also used as guidance on how to use \acvprust{} to discover bugs in a cryptographic library.

\subsection{Mozilla {NSS}}

NSS is an open-source library, maintained by Mozilla, Red Hat and others, providing support for a
variety of security standards. The standards include network security protocols (such as {TLSv1.3}), as well as many cryptography standards (like {PKCS\#3} and {PKCS\#11}). The library implements the most common cryptographic algorithms used in both public-key and private-key cryptography.
The NSS library is used by a huge variety of products, including the following: Mozilla products (including Firefox and Thunderbird), open-source client applications (such as LibreOffice), Red Hat and Oracle server products, and others. Security researchers repeatedly ran tests and targeted the library, which is covered under several bug bounty programs.

Since the library implements security-critical functionality, the code has also been extensively tested. Further, we will be exclusively referencing
testing applied to the cryptographic primitives and not the network protocols, parsers or other elements of the library.

Any modification occurring
to in the library has to pass through Continuous Integration tests. All these tests for cryptographic primitives could be divided into two big groups.
The first group of tests uses an internal application called \href{https://www.rdocumentation.org/packages/exomePeak/versions/2.6.0/topics/bltest}{\textbf{bltest}}.
The application allows to quickly check that the modifications in the code do not break the cryptography. For each primitive implemented in NSS, \texttt{bltest} provides several test vectors, provided by {NIST}.

\begin{enumerate}[\bfseries 1.]
    \item \href{https://searchfox.org/nss/source/cmd/bltest/tests/sha256}{SHA2}: 8 tests;
    \item \href{https://searchfox.org/nss/source/cmd/bltest/tests/aes_gcm}{AES-GCM}: 18 tests. The tests come from the original paper describing the standard~\cite{aes-gcm};
    \item \href{https://searchfox.org/nss/source/cmd/bltest/tests/rsa}{RSA}/\href{https://searchfox.org/nss/source/cmd/bltest/tests/rsa_pss}{PSS}/\href{https://searchfox.org/nss/source/cmd/bltest/tests/rsa_oaep}{OAEP}: 1/18/22 tests. The latter has {SHA1} and {SHA2} variants;
    \item \href{https://searchfox.org/nss/source/cmd/bltest/tests/ecdsa}{ECDSA}: 21 tests.
\end{enumerate}

The files in \texttt{bltest} contain test vectors for {ECDSA} using the {NIST} {P-256} curve (test vectors from~0 to~6 included), using the {NIST} {P-384} curve (test vectors from~7 to~13 included), and using the {NIST} {P-521} curve (test vectors from~14 to~20 included).

As the number of test vectors in \texttt{bltest} is limited, a second group of additional tests is performed each time the code in NSS changes, implemented using Google \texttt{gtests}~\cite{gtest} facility.
These tests (together with the \texttt{wycheproof}~\cite{wycheproof} tests run as a part of \texttt{gtests}), allow the developers to gain deeper confidence in the code.
\texttt{Wycheproof} tests include, among others, {AES-GCM}, {ECDSA} {P-256}, {P-384}, {P-521}, and {RSA}, which are also implemented in the current \acvprust{} NSS runner.

As more cryptographic functions are implemented using formal verification, the library
relies less on testing. However, formally verified code is still covered by
constant-time tests and fuzzed corpuses.

\subsection{Improving {NSS} Testing Coverage with \acvprust{}}

As part of a project to improve {NSS} testing infrastructure, we have developed an {NSS} runner for \acvprust{} and some extensions to the ACVP standard to cover more code. Specifically, we added a
\code{private\_key} structure to {RSA} and {ECDSA} test cases to allow the test case to specify the key
when generating the signature,
and implemented a \code{bn} (big number) subspecification that tests
big numbers directly, avoiding the lack of deep coverage that is resulting from testing higher-level API.
NSS runner supports most of the \code{sha}, \code{symmetric}, \code{rsa}, \code{ecdsa} published {ACVP} subspecifications.

\subsection{RSA Modulus Bug}

While working on {RSA} coverage with \acvprust{}'s {NSS} runner, we have discovered the following issue.
We describe the issue and the fix here to illustrate the methodology of discovering bugs using \acvprust{}.

{NSS} functions implementing {RSA} operations call for a couple of similar functions \code{rsa\_modulusLen} and \code{rsa\_modulusBits} to strip leading zeroes from
modulus bytes (see \autoref{listing:modulus_bugs}).

\begin{figure}[!ht]
\begin{minted}{c}
static unsigned int
rsa_modulusLen(SECItem *modulus)
{
    unsigned char byteZero = modulus->data[0];
    unsigned int modLen = modulus->len - !byteZero;
    return modLen;
}
\end{minted}
\begin{minted}{c}
static unsigned int
rsa_modulusBits(SECItem *modulus)
{
    unsigned char byteZero = modulus->data[0];
    unsigned int numBits = (modulus->len - 1) * 8;

    if (byteZero == 0) {
        numBits -= 8;
        byteZero = modulus->data[1];
    }

    while (byteZero > 0) {
        numBits++;
        byteZero >>= 1;
    }

    return numBits;
}
\end{minted}
\caption{{NSS} functions determining the {RSA} modulus lengths, from \code{rsapkcs.c}}%
\label{listing:modulus_bugs}
\end{figure}

As
demonstrated in \autoref{listing:modulus_bugs},
they make assumptions about the length of the modulus and perform indexed array access before checking the array size. This may cause
access to \textit{unrelated} memory. As a result, decisions based on it
may lead to security issues. For example, the attacker can arrange for the next part of memory to contain data, the decision based on which will lead to falsely considering the signature being processed valid. The bug is reproducible using the public RSA API of NSS. \autoref{listing:rsa_bug_c} demonstrates how the bug can be triggered.

\begin{figure}[!ht]
\begin{minted}{c}
SECITEM_MakeItem(NULL, &key.publicExponent, "", 0);
SECITEM_MakeItem(NULL, &key.modulus, "", 0);

RSA_CheckSignPSS(&key, HASH_AlgSHA256,
  HASH_AlgSHA256, 0, NULL, 0, NULL, 0);
\end{minted}
\caption{Example code fragment triggering the memory issue in {RSA} modulus length check}%
\label{listing:rsa_bug_c}
\end{figure}

The bug is not exploitable via existing software using {NSS}, because an unrelated check for insecure key sizes in TLS code discards the problematic {RSA} keys before operations are performed on
them. However, a valid {ACVP} test case uses our extensions:
\autoref{listing:rsa_bug_json} causes improper memory access, thus increasing vulnerability for third party software using the {NSS} {RSA} interface directly.

\begin{figure}[!ht]
\begin{minted}{json}
{
  "algorithm": "RSA",
  "mode": "sigGen",
  "revision": "FIPS186-5",
  "testGroups": [{
    "hashAlg": "SHA2_384",
    "maskFunction": "mgf1",
    "modulo": 4096,
    "saltLen": 31,
    "sigType": "pss",
    "testType": "GDT",
    "tests": [{
      "message": "",
      "privateKey": {
        "coefficient" : "00",
        "exponent1" : "00",
        "exponent2" : "00",
        "modulus" : "00",
        "prime1" : "00",
        "prime2" : "00",
        "privateExponent" : "00",
        "publicExponent" : "00"
      },
      "tcId" : 0,
    }],
  }]
}
\end{minted}
\caption{ACVP test case triggering the RSA modulus check bug in NSS}%
\label{listing:rsa_bug_json}
\end{figure}

We submitted a fix for the bug that adds additional checks to ensure array index cannot be out of bounds using Mozilla's official bug tracker. The fix has been accepted by the maintainers and included in the next version of NSS.

This bug was \textit{not} caught because of lack of focus on abnormal inputs, despite {NSS} testing suite including {RSA} test vectors. This highlights both the need to include diverse test cases within the valid input limits in the test vectors as well as the effectiveness and usability of \acvprust{} in improving test coverage and identifying new vulnerabilities.

\subsection{Other Bugs}\label{subsec:OtherBugs}
Several other non-security-related issues have been discovered during {NSS} testing.
One example is parsing negative big numbers that was non-functional due to an apparent bug.
Such issues, while not leading to vulnerabilities or even inadvertently shielding from them, are still dangerous, because they obscure other bugs and interfere with code analysis.
Even if dealt with or worked around, other issues may arise.
\autoref{tab:BugsDiscovered} provides a list with all the bugs we discovered while using \acvprust.
``Issue'' is the short description of the issue, ``Security'' is whether the issue was deemed to be related to security, ``Fix submitted'' means we submitted a patch to Mozilla official bug tracker, ``Fix accepted'' means the patch was accepted by NSS maintainers and included in the next NSS version.

\begin{table}[ht!]
\caption{List of issues discovered in NSS}\label{tab:BugsDiscovered}
\centering
    \begin{tabular}{| m{16em} | c | c | c |}
        \hline
        \rowcolor{Gray}
        \textbf{Issue} & \textbf{Security} & \textbf{Fix Submitted} & \textbf{Fix Accepted} \\
        \hline
        Segmentation fault or buffer overflow when calling \code{RSA\_CheckSignPSS} with special input. & \cmark{} & \cmark{} & \cmark{} \\
        \hline
        Infinite loop in \code{RSA\_PopulatePrivateKey}. & \cmark{} & \cmark{} & \cmark{} \\
        \hline
        Fails to build with clang 15 due to set but not used variable. & \xmark{} & \xmark{} & \xmark{} \\
        \hline
        Fails to build with clang 15 and optimized build due to set but used only in an assert variable. & \xmark{} & \xmark{} & \xmark{} \\
        \hline
        Assertion failed with certain \code{mp\_exptmod} inputs. & \cmark{} & \cmark{} & \cmark{} \\
        \hline
        Negative sign on \code{mp\_int} is ignored when read from raw form. & \xmark{} & \cmark{} & \cmark{} \\
        \hline
        RSA overflow check fails with really small modulus. & \cmark{} & \cmark{} & \cmark{} \\
        \hline
    \end{tabular}

\end{table}

\subsection{Disclosure}\label{subsec:Disclosure}
All bugs discovered were securely disclosed to NSS maintainers and have since been fixed in the latest development version of the library. This serves as proof that \acvprust{} has the potential to significantly enhance the security of existing cryptographic libraries by improving the process of identifying and addressing previously undiscovered bugs.

\section{Analysis of Efficiency at Improving Coverage}\label{sec:AnalysisCover}

In this section, we elaborate on the efficiency of \acvprust{} at improving code coverage.
Usually, coverage of testing is measured in the percentage of covered code.
The quantity of code is measured in lines, functions, branches, or other important parts, depending on the testing level~\cite{block-coverage}.
Ultimately, the most important measure of testing is how many issues are prevented or discovered. The bugs we found in already tested code indicate that hybrid fuzzing reached the new code that required coverage.
To get an idea of how much coverage \acvprust{} generated tests provide, we took measurements of the coverage by the corpus generated by libFuzzer running on {NSS} code.
We used an \acvprust{} RSA mutator developed by hybrid fuzzing, for~1 hour, with a maximum input size of 10,000 bytes.
The experiments were performed on an Intel i7-12700 processor at 2100 MHz, using a single thread.
We used the current development version of {NSS} as of \code{Fri Sep 8 21:48:00 2023 +0000}.
To measure the efficiency, we consider coverage of the RSA code, its improvement over traditional coverage-guided fuzzing and over the existing {NSS} test suite.

\subsection{Scope of Generated Coverage}

Some areas of {RSA} code are excluded from coverage due to limitations of either {NSS} or the {ACVP} standard. Key generation and related code is excluded since {NSS} does \textit{not} provide an API for generating predictable keys. Additionally, ``block'' functions are excluded since their usage is mostly internal. Neither variant covers paths that involve running out of memory and other unexpected outside factors. The {ACVP} subspecification with our custom extensions covers:

\begin{enumerate}[\bfseries 1.]
  \item \textbf{Signature Generation and Verification:} PSS, {PKCS \#1} 1.5, and primitive modes, with multiple {SHA} variants as the digest function;
  \item \textbf{Encryption and Decryption:} {OAEP} and primitive modes;
  \item \textbf{Key Population:} As part of the above, missing private key components are generated from present ones.
\end{enumerate}

\subsection{Analysis of Improvement over Pure Coverage-Guided Fuzzing}

Coverage-guided fuzzing, such as employed by {libFuzzer}, is good at automatically covering most of the code, but it fails to satisfy particular criteria commonly present in public key cryptography implementations, thus omitting numerous potentially vulnerable code areas.

The most important RSA code is located in two source files, \code{rsa.c} and \code{rsapkcs.c}. The following list describes the remaining pieces of code not covered by modes of fuzzing and coverage differences between {libFuzzer} standard coverage guided fuzzing
and hybrid fuzzing enhanced by \acvprust.
The full coverage reports are available
\ifANON{}
in the anonymized source code repository: \href{https://anonymous.4open.science/api/repo/acvpframe/file/reports/rsa.insane.html}{RSA code coverage before using \acvprust{}} and after: \href{https://anonymous.4open.science/api/repo/acvpframe/file/reports/rsa.sane.html}{RSA code coverage after using \acvprust{}}.
\else{}
in the source code repository: \href{https://gitlab.com/nisec/acvp-rust/-/raw/master/reports/rsa.insane.html?inline=false}{RSA code coverage before using \acvprust{}} and after: \href{https://gitlab.com/nisec/acvp-rust/-/raw/master/reports/rsa.sane.html?inline=false}{RSA code coverage after using \acvprust{}}.
\fi{}
\begingroup{}
\raggedright{}\begin{enumerate}
  \item \code{rsapkcs.c:254:} Check for proper padding being present, leads to \code{rsaFormatBlock} never executed. Data size conditionally restricted.

  \item \code{DecryptOAEP\(\):} Coverage is missing from plain fuzzing due to key checks failing most of the time for fuzzing-generated keys. Restrictions added to make sure key components pass basic checks. This also causes \code{eme\_oaep\_decode} not to be covered in plain variant.

  \item \code{rsapkcs.c:1258 emsa\_pss\_encode:} Check for modulus length fails due to complicated relations between multiple lengths. Interlinked restrictions added on salt and modulus length to pass the check.

  \item \code{emsa\_pss\_verify:} Is not covered in plaintext versions due to \code{RSA\_PublicKeyOp} never succeeding in this context due to unmet conditions listed above.

  \item \code{rsapkcs.c:1669 RSA\_CheckSignRecover:} Hybrid version can pass the signature verification earlier, but further checks on decoded data fail. It is not feasible to improve
  coverage further.
\end{enumerate}
\endgroup{}
The end result is that using \acvprust{} helps the fuzzer to produce tests covering critical areas inaccessible by CGF with minimal human intervention.

\section{ACVP Test Vector Format}\label{sec:acvp}

As part of our design-related work, implementation and testing of \acvprust{} consisted of implementing the processes for both the parsing and handling of the ACVP test vector format.
During this process, we identified the need for improvement.
More precisely, we became aware of the need to render its implementation easier and safer in modern languages and improve the test transmission and storage efficiency.
In this section, we provide some suggestions on how to work towards achieving these improvements.
We propose to make the ACVP test format include more well-defined nested structures to make it more flexible and to make parsing easier.
We also suggest to make the tests simpler to write and combine by allowing user-controlled level of sharing data between groups of tests.

\subsection{Structures Usage}\label{subsec:StructUse}

In modern libraries parsing serialization formats, the parsing code is often generated from the declarative structure definition, like in \href{https://serde.rs/derive.html}{Serde}.
This approach produces safe code with automatic error handling.
In ACVP subspecifications structures like encryption keys are included in the parent structure as a set of optional fields.
All or none of these fields should be present, but such check has to be written manually.
Additionally, these combinations of fields are often repeated.
Moving them to a separate structure could improve readability and maintainability of the specification as well as its implementations, as can be seen in \autoref{listing:rsa-flatten} vs \autoref{listing:rsa-separate}.

\begin{figure}[!ht]
\begin{minted}{c}
{
	"d" : "02",,
	"message" : "ffffff21ff",
	"q" : "00ffffffffffff21ff",
	"dmp1" : "00",
	"tcId" : 0,
}
\end{minted}
\caption{Example of an RSA ACVP test case with private key flattened into main structure}%
\label{listing:rsa-flatten}
\end{figure}

\begin{figure}[!ht]
\begin{minted}{c}
{
	"message" : "ffffff21ff",
	"privateKey" : {
		"coefficient" : "00",
		"prime2" : "00ffffffffffff21ff",
		"privateExponent" : "02",
	},
	"tcId" : 0,
}
\end{minted}
\caption{Example of an RSA ACVP test case with private key separated into a structure}%
\label{listing:rsa-separate}
\end{figure}

\subsection{Level-Specific Fields}

ACVP vector sets include three levels: Test Vector, Test Group, and Test Case.
Each of them can contain a combination of multiple sublevels.
Some levels may include fields that affect lower levels.
This was clearly intended as a simplification measure, but in case users need to implement multiple test cases with different attributes only available at the higher level, the complexity of the vector set actually grows, as multiple vector sets or test vectors need to be introduced.
To remedy that, we propose rendering said fields universal by providing the option of adding both at test case and test group level and making test groups recursive, so that test cases may be grouped in a flexible manner, as can be seen in \autoref{listing:rsa-repeated-fields} vs \autoref{listing:rsa-separate-fields}.

\begin{figure}[!ht]
\begin{minted}{c}
"testType" : "GDT",
"tests" : [
	{
		"d" : "fefff",
		"message" : "",
		"n" : "fd12",
		"p" : "136",
		"q" : "1",
		"tcId" : 0
	},
	{
		"d" : "fefff",
		"message" : "",
		"n" : "fd12",
		"p" : "ff",
		"q" : "1254",
		"tcId" : 3
	},
	{
		"d" : "fefff",
		"message" : "",
		"n" : "fd12",
		"p" : "ff",
		"q" : "36fa",
		"tcId" : 6
	},
]
\end{minted}
\caption{Example of an RSA ACVP test group with fields repeated for every test case}%
\label{listing:rsa-repeated-fields}
\end{figure}

\begin{figure}[!ht]
\begin{minted}{c}
"testType" : "GDT",
"testFields": {
	"d" : "fefff",
	"message" : "",
},
"tests" : [
	{
		"p" : "136",
		"q" : "1",
		"tcId" : 0
	},
	{
		"p" : "ff",
		"q" : "1254",
		"tcId" : 3
	},
	{
		"p" : "ff",
		"q" : "36fa",
		"tcId" : 6
	},
]
\end{minted}
\caption{Example of an RSA ACVP test group with shared fields in one place}%
\label{listing:rsa-separate-fields}
\end{figure}

\section{Conclusion}\label{sec:conclusion}

In this paper, we presented \acvprust{} -- a software framework for analyzing cryptographic libraries, whose main aim is to discover possible bugs in the code. Through a series of experiments, we have demonstrated that \acvprust{} produces efficient covering tests that can be shared between cryptographic libraries. Furthermore, it provides a base that facilitates the structure of an adaptor for a new library. In addition, it creates sets of tests that not only increase confidence about how correct implemented algorithms are, but also provides good coverage, capable of using knowledge gained from researches conducted in other libraries.

Additionally, we used \acvprust{} to analyze Mozilla's {NSS} cryptographic library. This allowed us to trace new, undiscovered bugs in this widely-used library. The identified bugs have been disclosed and accepted by maintainers. This serves as proof that \acvprust{} facilitates the detection of undiscovered bugs and has the potential to improve the security of existing software with a main focus on cryptographic libraries. Furthermore, we showed that \acvprust{} increases code coverage compared to other tools. This leads to significant improvements in fuzzing quality and helps to detect issues in otherwise hard-to-reach code areas. Finally, in order to support open science and reproducible research, we have made \acvprust{} publicly available.


Experience has shown that it is important to include diverse test cases in test suites to ensure both
corner cases are not missing, and code hidden behind complex conditions is covered. This is what \acvprust{} methodology allows a researcher to do.

\subsection{Future Research}\label{subsec:Future}
Future possibilities for improving the work include development of further subspecifications, with the goal of providing more input flexibility to increase coverage,
possible further automation of the process, and automatically discovering side-channel vulnerabilities by integrating related tools.



\ifACM{}
\bibliographystyle{ACM-Reference-Format}
\fi

\ifLNCS{}
\bibliographystyle{splncs04}
\fi

\ifIEEE{}
\bibliographystyle{IEEEtranN}
\fi

\ifUSENIX{}
\bibliographystyle{plainnat}
\fi

\ifIACRTRANS{}
\printbibliography{}
\else
\bibliography{manuscript_rw,manuscript_ro}
\fi

\end{document}